\documentclass[titlepage,letterpaper,12pt]{article}
\usepackage[dvips]{graphicx}

\begin{document}

\title{Exact solution for a fermion in the background of a scalar inversely linear
potential}
\date{}
\author{Antonio S. de Castro \\
%EndAName
\\
UNESP - Campus de Guaratinguet\'{a}\\
Departamento de F\'{\i}sica e Qu\'{\i}mica\\
Caixa Postal 205\\
12516-410 Guaratinguet\'{a} SP - Brasil\\
\\
E-mail address: castro@feg.unesp.br (A.S. de Castro)}
\date{}
\maketitle

\begin{abstract}
The problem of a fermion subject to a general scalar potential in a
two-dimensional world is mapped into a Sturm-Liouville problem for nonzero
eigenenergies. The searching for possible bounded solutions is done in the
circumstance of power-law potentials. The normalizable zero-eigenmode
solutions are also searched. For the specific case of an inversely linear
potential, which gives rise to an effective Kratzer potential, exact bounded
solutions are found in closed form. The behaviour of the upper and lower
components of the Dirac spinor is discussed in detail and some unusual
results are revealed.
\end{abstract}

The Coulomb potential of a point electric charge in a 1+1 dimension,
considered as the time component of a Lorentz vector, is linear ($\sim |x|$)
and so it provides a constant electric field always pointing to, or from,
the point charge. This problem is related to the confinement of fermions in
the Schwinger and in the massive Schwinger models \cite{col1}-\cite{col2}
and in the Thirring-Schwinger model \cite{fro}. It is frustrating that, due
to the tunneling effect (Klein\'{}s paradox), there are no bound states for
this kind of potential regardless of the strength of the potential \cite{cap}%
-\cite{gal}. The linear potential, considered as a Lorentz scalar, is also
related to the quarkonium model in one-plus-one dimensions \cite{hoo}-\cite
{kog}. Recently it was incorrectly concluded that even in this case there is
solely one bound state \cite{bha}. Later, the proper solutions for this last
problem were found \cite{cas1}-\cite{hil}. The mixed vector-scalar potential
has also been analyzed for a linear potential \cite{cas2} as well as for a
general potential which goes to infinity as $|x|\rightarrow \infty $ \cite
{ntd}. In both of those last references it has been concluded that there is
confinement if the scalar coupling is of sufficient intensity compared to
the vector coupling.

The problem of a particle subject to an inversely linear potential in one
spatial dimension ($\sim |x|^{-1}$), known as the one-dimensional hydrogen
atom, has received considerable attention in the literature. This problem
presents some conundrums and the most perplexing is that one regarding the
ground state. The nonrelativistic Schr\"{o}dinger equation provides a
ground-state solution with infinite eigenenergy and a related eigenfunction
given by a delta function centered about the origin. This problem was also
analyzed with the Klein-Gordon equation and there it was revealed a finite
eigenenergy and a exponentially decreasing eigenfunction \cite{spe}. By
using the technique of continuous dimensionality the problem was approached
with the Schr\"{o}dinger, Klein-Gordon and Dirac equations. The conclusion
in this more recent work is that the Klein-Gordon equation provides
unacceptable solutions while the Dirac equation, with the interacting
potential considered as a time component of a vector, has no bounded
solutions at all \cite{mos}. This problem was also sketched for a Lorentz
scalar interacting potential in the Dirac equation \cite{ho}, but  the
analysis is incomplete.

In the present paper it is shown that the problem of a fermion under the
influence of a general scalar potential for nonzero eigenenergies can be
mapped into a Sturm-Liouville problem. Next the key conditions for the
existence of bound-state solutions are settled for power-law potentials. The
possible zero-mode solutions are shown to conform with the ultrarelativistic
limit of the theory. In addition, the solution for an inversely linear
potential is obtained in closed form. The effective potential resulting from
the mapping has the form of the Kratzer potential \cite{kra}. It is
noticeable that this problem has an infinite number of acceptable bounded
solutions, nevertheless it has no nonrelativistic limit for small quantum
numbers. It is also shown that in the regime of strong coupling additional
zero-energy solutions can be obtained as a limit case of nonzero-energy
solutions. The ideas of supersymmetry have already been used to explore the
two-dimensional Dirac equation with a scalar potential \cite{coo}-\cite{nog}%
, nevertheless the power-law potential has been excluded of such discussions.

The two-dimensional Dirac equation can be obtained from the
four-dimen\-sional one with the mixture of spherically symmetric scalar,
vector and anomalous magnetic-like (tensor) interactions. If we limit the
fermion to move in the $x$-direction ($p_{y}=p_{z}=0$) the four-dimensional
Dirac equation decomposes into two equivalent two-dimensional equations with
2-component spinors and 2$\times $2 matrices \cite{str}. Then, there results
that the scalar and vector interactions preserve their Lorentz structures
whereas the anomalous magnetic interaction turns out to be a pseudoscalar
interaction. Furthermore, in the 1+1 world there is no angular momentum so
that the spin is absent. Therefore, the 1+1 dimensional Dirac equation allow
us to explore the physical consequences of the negative-energy states in a
mathematically simpler and more physically transparent way.

In the presence of a time-independent scalar potential the 1+1 dimensional
time-independent Dirac equation for a fermion of rest mass $m$ reads

\begin{equation}
\left[ c\alpha p+\beta \left( mc^{2}+V\right) \right] \Psi =E\Psi  \label{1}
\end{equation}

\noindent where $E$ is the energy of the fermion, $c$ is the velocity of
light and $p$ is the momentum operator. $\alpha $ and $\beta $ are Hermitian
square matrices satisfying the relations $\alpha ^{2}=\beta ^{2}=1$, $%
\left\{ \alpha ,\beta \right\} =0$. From the last two relations it follows
that both $\alpha $ and $\beta $ are traceless and have eigenvalues equal to
$\pm $1, so that one can conclude that $\alpha $ and $\beta $ are
even-dimensional matrices. One can choose the 2$\times $2 Pauli matrices
satisfying the same algebra as $\alpha $ and $\beta $, resulting in a
2-component spinor $\Psi $. The positive definite function $|\Psi |^{2}=\Psi
^{\dagger }\Psi $, satisfying a continuity equation, is interpreted as a
probability position density and its norm is a constant of motion. This
interpretation is completely satisfactory for single-particle states \cite
{tha}. Using $\alpha =\sigma _{2}$, $\beta =\sigma _{1}$ and provided that
the spinor is written in terms of the upper and the lower components
\begin{equation}
\Psi =\left(
\begin{array}{c}
\Psi _{+} \\
\Psi _{-}
\end{array}
\right)  \label{2}
\end{equation}

\noindent the Dirac equation decomposes into :

\begin{eqnarray}
E\Psi _{+} &=&\left( mc^{2}+V\right) \Psi _{-}-\hbar c\Psi _{-}^{\prime }
\nonumber \\
&&  \label{3} \\
E\Psi _{-} &=&\left( mc^{2}+V\right) \Psi _{+}+\hbar c\Psi _{+}^{\prime }
\nonumber
\end{eqnarray}

\noindent where the prime denotes differentiation with respect to $x$. In
terms of $\Psi _{+}$ and $\Psi _{-}$ the spinor is normalized as $%
\int_{-\infty }^{+\infty }dx\left( |\Psi _{+}|^{2}+|\Psi _{-}|^{2}\right) =1$%
, so that $\Psi _{+}$ and $\Psi _{-}$ are square integrable functions. It is
remarkable that the Dirac equation with a scalar potential is not invariant
under $V\rightarrow V+const.$ Therefore, the absolute values of the energy
will have physical significance and the freedom to choose a zero-energy will
be lost.

The coupling between the upper and the lower components of the Dirac spinor
can be formally eliminated for $E\neq 0$ when Eq. (\ref{3}) is written a
pair of second-order differential equations

\begin{equation}
-\frac{\hbar ^{2}}{2m}\;\Psi _{\pm }^{\prime \prime }+V_{eff}^{\pm }\;\Psi
_{\pm }=E_{eff}\;\Psi _{\pm }  \label{8}
\end{equation}

\noindent where
\begin{eqnarray}
E_{eff} &=&\frac{E^{2}-m^{2}c^{4}}{2mc^{2}}  \label{9} \\
&&  \nonumber \\
V_{eff}^{\pm } &=&\frac{V^{2}}{2mc^{2}}+V\mp \frac{\hbar }{2mc}V^{\prime }
\label{10}
\end{eqnarray}

\bigskip \noindent \noindent \noindent \noindent These last results tell us
that the solution for this class of problem consists in searching for
bounded solutions for two Schr\"{o}dinger equations. It should not be
forgotten, though, that the equations for $\Psi _{+}$ or $\Psi _{-}$ are not
indeed independent because the effective eigenvalue $E_{eff}$, appears in
both equations. Therefore, one has to search for bound-state solutions for $%
V_{eff}^{+}$ and $V_{eff}^{-}$ with a common eigenvalue. The Dirac
eigenvalues are obtained by inserting the effective eigenvalues in (\ref{9}).

One should realize that the Dirac energy levels are symmetrical about $E=0$
(see, \textit{e.g.}, Refs. \cite{cn} and \cite{cnt}). This conclusion can be
obtained directly from (\ref{9}) as well as from the charge conjugation.
Indeed, if $\Psi $ is a solution with energy $E$ then $\sigma _{3}\Psi ^{*}$
is also a solution with energy $-E$ for the very same potential. It means
that the potential couples to the positive-energy component of the spinor in
the same way it couples to the negative-energy component. In other words,
this sort of potential couples to the mass of the fermion instead of its
charge so that there is no atmosphere for the spontaneous production of
particle-antiparticle pairs. No matter the intensity and sign of the
coupling parameter, the positive- and the negative-energy solutions never
meet. Thus there is no room for transitions from positive- to
negative-energy solutions. This all means that Klein\'{}s paradox never
comes to the scenario.

It is worth to note that the Dirac equation is covariant under $x\rightarrow
-x$ if $V(x)$ does not change sign and $\Psi_\pm (-x)=\Psi_\mp (x)$ or $%
\Psi_\pm (-x)=-\Psi_\mp (x)$. This is because the parity operator $P=\exp
(i\eta )P_{0}\sigma _{1}$, where $\eta $ is a constant phase and $P_{0}$
changes $x$ into $-x$, changes sign of $\alpha $ but not of $\beta $. For an
even-parity potential one can notice that $V_{eff}^{\pm }(-x)=V_{eff}^{\mp
}(x)$ whereas $E_{eff}$ remains unchanged.

Now let us consider a scalar potential in the form in the form $V=\mu
|x|^{\delta }$, then the effective potential becomes

\begin{equation}
V_{eff}^{\pm }=\frac{\mu ^{2}}{2mc^{2}}|x|^{2\delta }+\mu |x|^{\delta }\mp
\varepsilon (x)\frac{\hbar \mu \delta }{2mc}|x|^{\delta -1}  \label{11}
\end{equation}

\noindent where $\varepsilon (x)$ stands for the sign function. In the
subsequent survey for the possible existence of bounded solutions we take
advantage of the symmetry for $V_{eff}^{\pm }$ and only consider the
positive side of the $x$-axis. When $\delta >0$ the effective potential goes
to infinity as $x\rightarrow \infty $ and it is finite at the origin for $%
\delta \geq 1$ whereas in the range $0<\delta <1$ it  has a singularity
given by $\mp \hbar \mu \delta /2mc|x|^{1-\delta }$, implying in a
potential-well structure for $V_{eff}^{-}$ ($V_{eff}^{+}$) when $\mu >0$ ($%
\mu <0$) and an attractive potential less singular than $x^{-1}$ for $%
V_{eff}^{+}$ ($V_{eff}^{-}$). Therefore, for $\delta >0$ the power potential
leads to effective potentials fulfilling the key conditions to furnish
discrete spectra. On the other hand, when $\delta <0$ the effective
potential vanishes as $x\rightarrow \infty $ and when $\mu >0$ the effective
potential $V_{eff}^{+}$ is always repulsive at the origin in such a way that
it is repulsive everywhere. Therefore, for $\delta <0$ and $\mu >0$ the
power-law potential does not lead to bound-state solutions. For $\delta <0$
and $\mu <0$, though, there is a potential-well structure for $V_{eff}^{-}$.
The same is true for $V_{eff}^{+}$ on the condition that $\delta <-1$. For $%
-1<\delta <0$ and $\mu <0$ there is an attractive potential $V_{eff}^{+}$
\thinspace with singularity given by $-\hbar |\mu ||\delta
|/2mc|x|^{1+|\delta |}$. For $\delta =-1$ and $\mu <0$ there is also a
potential-well structure for $V_{eff}^{+}$ when $\mu <-\hbar c$, and an
attractive potential $V_{eff}^{+}$ with singularity at the origin given by $%
-|\mu |/|x|$ when $\mu =-\hbar c$, and an attractive
inverse-square singularity given by $-\hbar |\mu |\left( 1-|\mu
|/\hbar c\right) /2mcx^{2}$ when $-\hbar c<\mu <0$. There is no
collapse to the center because the potential is not more
attractive than $-\hbar ^{2}/8mx^{2}$ \cite{lan}. As a matter of
fact, when  $\mu =-\hbar c/2$ the singular potential assumes  its
critical value. Therefore, for $\delta <0$ and $\mu <0$ the
power-law potential also leads to effective potentials fulfilling
the key conditions to furnish discrete spectra.

Up to this point we have considered solutions for $E\neq 0$. This
supposition has been explicitly assumed to obtain (\ref{8})-(\ref{10}).
Nevertheless, one could also ask for possible zero-energy solutions. These
zero-mode energies can be obtained directly from the Dirac equation (\ref{3}%
). In this case the first-order differential equations are fully uncoupled.
Then the upper and lower components of the Dirac spinor for the power-law
potential are expressed by

\begin{equation}
\Psi _{\pm }=N_{\pm }F_{\pm }(x)\exp \left( \mp \frac{mc}{\hbar }x\right)
\label{28}
\end{equation}

\noindent where
\begin{equation}
F_{\pm }(x)=\left\{
\begin{array}{c}
|x|^{\mp \varepsilon (x)\frac{\mu }{\hbar c}} \\
\\
\exp \left( \mp \frac{\mu }{\hbar c}\varepsilon (x)\frac{|x|^{\delta +1}}{%
\delta +1}\right)
\end{array}
\begin{array}{l}
,\hspace{0.25cm}\delta =-1 \\
\\
,\hspace{0.25cm}{\rm{for}}\hspace{0.25cm}\delta \neq -1
\end{array}
\right.  \label{28a}
\end{equation}
and $N_{\pm }$ is a normalization constant. Normalizable eigenspinors on the
positive side of the $x$-axis are given by
\begin{equation}
\Psi =\left\{
\begin{array}{c}
\Psi _{+}\left(
\begin{array}{l}
1 \\
0
\end{array}
\right) \\
\\
\Psi _{-}\left(
\begin{array}{l}
0 \\
1
\end{array}
\right)
\end{array}
\begin{array}{l}
,\hspace{0.25cm}{\rm{for}}\hspace{0.25cm}\mu >0\hspace{0.25cm}{\rm{and}}%
\hspace{0.25cm}\ \delta
>-1\hspace{0.25cm}{\rm{or}}\hspace{0.25cm}\mu <0\
\hspace{0.25cm}{\rm{and}}\hspace{0.25cm}\delta \leq -1 \\
\\
,\hspace{0.25cm}{\rm{for}}\hspace{0.25cm}\mu <0\ \hspace{0.25cm}{\rm{and}%
}\hspace{0.25cm}\delta >-1
\end{array}
\right.  \label{28b}
\end{equation}

\noindent Note that the probability position density of the zero-mode spinor
has a lonely hump. Note also that the conditions for the existence of
bounded solutions for zero-energies has nothing to do with those ones for $%
E\neq 0$. In the case of $\delta =-1$ another restriction must be added to
the top line of (\ref{28b}): $\mu \leq -\hbar c$. This restriction is
necessary for obtaining a differentiable spinor at the origin and it means
that the inversely linear potential must be enough strong to hold a
zero-mode solution.

It is surprising to find Dirac eigenspinors with a vanishing lower component
in a theory without a nonrelativistic limit. More surprising is to find a
vanishing upper component. Both dramatic circumstances make their appearance
due to the particular representations of the matrices $\alpha $ and $\beta $
adopted in this paper. It is instructive at this point to consider for a
moment a representation where the eigenspinor presents a more familiar
behaviour. Let us write the Dirac equation (\ref{1}) as

\bigskip
\begin{equation}
\left[ c\sigma _{1}p+\sigma _{3}\left( mc^{2}+V\right) \right] \tilde{\psi}=E%
\tilde{\psi}  \label{29}
\end{equation}

\noindent The original spinor is related to $\tilde{\psi}$ by the unitary
transformation $\psi =U\tilde{\psi}$, where

\begin{equation}
U=\frac{1}{\sqrt{2}}\left(
\begin{array}{ll}
1 & -i \\
1 & +i
\end{array}
\right)  \label{30}
\end{equation}

\noindent so that $\phi =\left( \tilde{\phi}-i\tilde{\chi}\right) /\sqrt{2}$
and $\chi =\left( \tilde{\phi}+i\tilde{\chi}\right) /\sqrt{2}$. In the
nonrelativistic approximation (\ref{29}) becomes

\begin{equation}
\tilde{\chi}=\frac{p}{2mc}\;\tilde{\phi}  \label{eq8c}
\end{equation}

\begin{equation}
\left( -\frac{\hbar ^{2}}{2m}\frac{d^{2}}{dx^{2}}+V\right) \tilde{\phi}%
=\left( E-mc^{2}\right) \tilde{\phi}  \label{eq8d}
\end{equation}

\noindent Eq. (\ref{eq8c}) shows that $\tilde{\chi}$ is of order $v/c<<1$
relative to $\tilde{\phi}$ and Eq. (\ref{eq8d}) shows that $\tilde{\phi}$
obeys the Schr\"{o}dinger equation with the potential $V$. Now one can see
that when one uses the representation where $\alpha =\sigma _{2}$, $\beta
=\sigma _{1}$ (that one used in this paper) one obtains upper and lower
components approximately equal to each other in the nonrelativistic limit.
On the other side, in the ultrarelativistic limit one expects that $\tilde{%
\chi}$ presents a contribution comparable to $\tilde{\phi}$, thus the
possibilities $\tilde{\phi}\approx i\tilde{\chi}$ and $\tilde{\phi}\approx -i%
\tilde{\chi}$ imply into $\phi \approx 0$ and $\chi \approx 0$,
respectively. Therefore, one can conclude that the zero-mode solutions given
by (\ref{28b}) correspond to the ultrarelativistic limit of the
theory.\noindent \bigskip

Now let us focus our attention on a scalar power-law potential in the form
\begin{equation}
V=-\frac{\hbar cq}{|x|}  \label{12}
\end{equation}
\noindent where $q$ is a real parameter. Then the effective potential
becomes the Kratzer potential
\begin{equation}
V_{eff}^{\pm }=-\frac{\hbar cq}{|x|}+\frac{A_{\pm }}{x^{2}}  \label{13}
\end{equation}
\noindent where
\begin{equation}
A_{\pm }=\frac{\hbar ^{2}q}{2m}\left[ q\mp \varepsilon (x)\right]  \label{14}
\end{equation}

\begin{figure}[!ht]
\begin{center}
\includegraphics[width=8cm, angle=270]{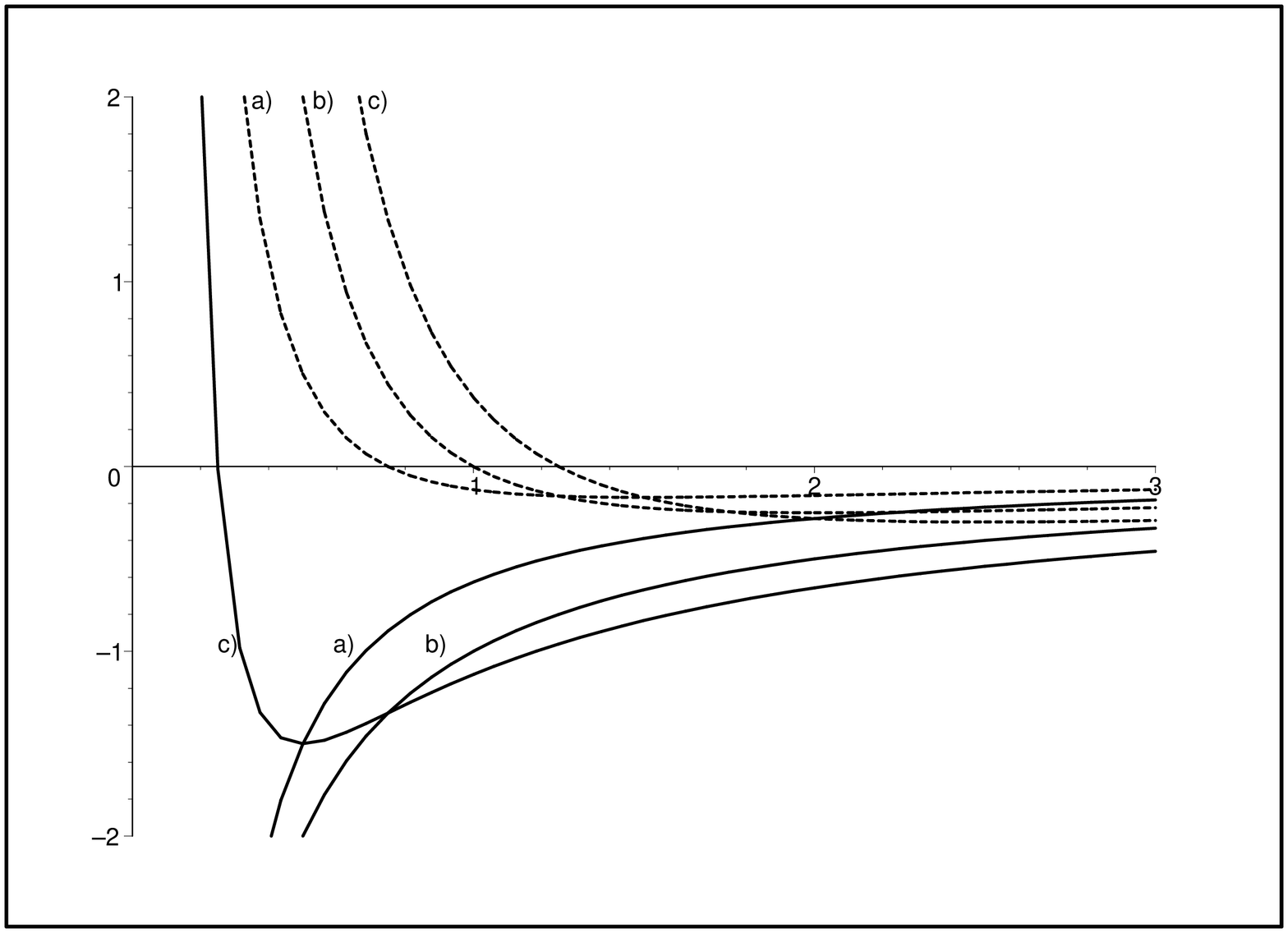}
\end{center}
\par
\vspace*{-0.1cm}
\caption{Effective Kratzer potentials as functions of $x$. The solid lines
are for $V_{eff}^{+}$, the dashed lines are for $V_{eff}^{-}$. \quad a) $%
q=1/2\,$; \quad b) $q=1\,$; \quad c) $q=3/2$ \quad ($m=\hbar =c=1$).}
\label{Fig1}
\end{figure}

\noindent As seen in the preceding paragraph this potential is able to bind
fermions on the condition that $q>0$. The effective Kratzer potential is
plotted in Fig. \ref{Fig1} for some illustrative values of $q$. It follows
that $E_{eff}<0$, corresponding to Dirac eigenvalues in the range $%
-mc^{2}<E<+mc^{2}$. The Schr\"{o}dinger equation with the Kratzer potential
is an exactly solvable problem and its solution, for a repulsive
inverse-square term in the potential ($A_{\pm }>0$), can be found on
textbooks \cite{lan}-\cite{flu}. Since we need solutions involving a
repulsive as well as an attractive inverse-square term in the potential, the
calculation including this generalization is presented.

Defining the dimensionless quantities $\xi $ and $C$,

\negthinspace
\begin{equation}
\xi =\frac{2}{\hbar }\sqrt{-2mE_{eff}}\;|x|\qquad \textrm{and}\mathrm{\qquad }%
C=q\sqrt{-\frac{mc^{2}}{2E_{eff}}}  \label{15}
\end{equation}

\noindent and using (\ref{8})-(\ref{9}) and (\ref{13}) one obtains the
equation

\begin{equation}
\Psi _{\pm }^{\prime \prime }+\left( -\frac{1}{4}+\frac{C}{\xi }-\frac{%
2mA_{\pm }}{\hbar ^{2}\xi ^{2}}\right) \Psi _{\pm }=0  \label{16}
\end{equation}

\noindent Now the prime denotes differentiation with respect to $\xi $. The
normalizable asymptotic form of the solution as $\xi \rightarrow \infty $ is
$e^{-\xi /2}$. As $\xi \rightarrow 0$, when the term $1/x^{2}$ dominates,
the regular solution behaves as $\xi ^{s}$, where $s$ is a nonnegative
solution of the algebraic equation

\begin{equation}
s(s-1)-\frac{2mA_{\pm }}{\hbar ^{2}}=0  \label{17}
\end{equation}
\textit{viz.}

\begin{equation}
s=\frac{1}{2}\left( 1\pm \sqrt{1+\frac{8mA_{\pm }}{\hbar ^{2}}}\right) \geq 0
\label{18}
\end{equation}

\noindent If $A_{\pm }>0$ there is just one possible value for $s$ (that one
with the plus sign in front of the radical) and the same is true for $A_{\pm
}=A_{c}=-\hbar ^{2}/8m$ when $s=1/2$, but for $A_{c}<A_{\pm }<0$ there are
two possible values for $s$ in the interval $0<s<1$. If the inverse-square
potential is absent ($A_{\pm }=0$) then $s=0$ or $s=1$. The solution for all
$\xi $ can be expressed as $\Psi _{\pm }(\xi )=\xi ^{s}e^{-\xi /2}w(\xi )$,
where $w$ is solution of the confluent hypergeometric equation \cite{abr}

\begin{equation}
\xi w^{\prime \prime }+(b-\xi )w^{\prime }-aw=0  \label{19}
\end{equation}

\noindent with

\begin{eqnarray}
a &=&\frac{b}{2}-C  \nonumber \\
&&  \label{20} \\
b &=&2s  \nonumber
\end{eqnarray}

\noindent Then $w$ is expressed as $_{1\!\;}\!F_{1}(a,b,\xi )$ and in order
to furnish normalizable $\psi _{\pm }$, the confluent hypergeometric
function must be a polynomial. This demands that $a=-N$, where $N$ is a
nonnegative integer in such a way that $_{1\!\;}\!F_{1}(a,b,\xi )$ is
proportional to the associated Laguerre polynomial $L_{N}^{b-1}(\xi )$, a
polynomial of degree $N$. This requirement, combined with the top line of (%
\ref{20}), also implies into quantized effective eigenvalues:

\begin{equation}
E_{eff}=-\frac{q^{2}}{2\left( s+N\right) ^{2}}\;mc^{2},\qquad N=0,1,2,\ldots
\label{21}
\end{equation}

\noindent with eigenfunctions given by

\begin{equation}
\Psi _{\pm }(\xi )=N_{\pm }\;\xi ^{s}e_{\;}^{-\xi /2}\;L_{N}^{2s-1}\left(
\xi \right) ,\qquad s>0  \label{22}
\end{equation}

\smallskip

\noindent

\noindent where the new constraint over $s$ is a consequence of the
definition of associated Laguerre polynomials. Note that the behavior of $%
\Psi _{\pm }$ at very small $\xi $ implies into the Dirichlet boundary
condition ($\Psi _{\pm }(0)=0)$. This boundary condition is essential
whenever $A_{\pm }\neq 0$, nevertheless it also develops for $A_{\pm }=0$.

The necessary conditions for binding fermions in the Dirac equation with the
effective Kratzer potential have been put forward. The formal analytical
solutions have also been obtained. Now we move on to consider a survey for
distinct cases in order to match the common effective eigenvalue. As we will
see this survey leads to additional restrictions on the solutions, including
constraints involving the nodal structure of the upper and lower components
of the Dirac spinor.

\begin{figure}[!ht]
\begin{center}
\includegraphics[width=8cm, angle=270]{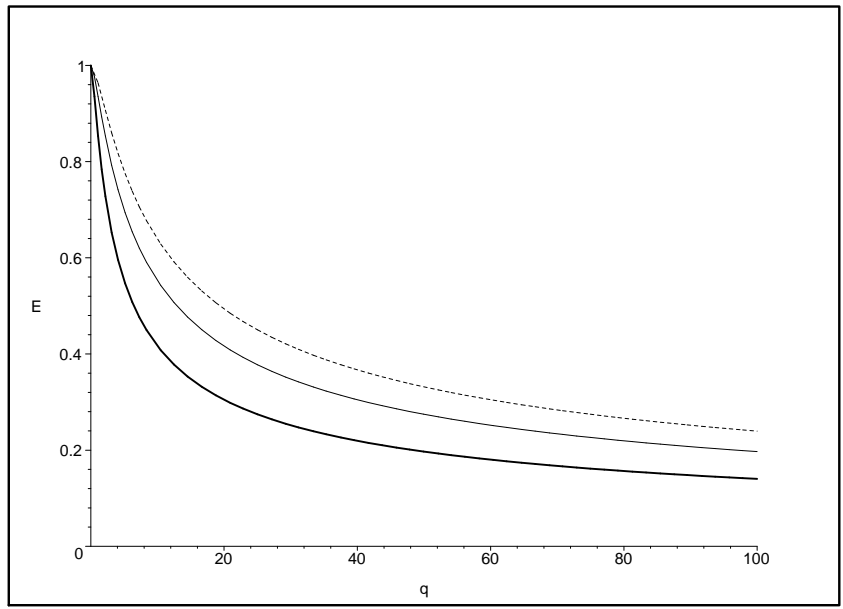}
\end{center}
\par
\vspace*{-0.1cm} \caption{The three lowest positive eigenvalues
($E$) as a function of $q$  for the potential $V=-\hbar cq/|x|$
($m=c=1$). The full thick line
stands for $n=1$, the full thin line for $n=2$ and the dashed line for $n=3$%
. }
\label{Fig2}
\end{figure}

When\textbf{\ }$q=1/2$ one has $A_{-}>0$ with $s=3/2$, and $A_{+}=A_{c}<0$
with $s=1/2$. For\textbf{\ }$q=1$, $A_{-}>0$ with $s=2$, and $A_{+}=0$ with $%
s=$ $1$. For $q>1$ one has $s=q+(1{\mp }1)/2$, corresponding to $A_{\pm }>0$%
. For $0<q<1$ and $q\neq 1/2$, though, one has $A_{-}>0$ with $s=q+1$ and $%
A_{c}<A_{+}<0$ with $s=q$ or $s=-q+1$. The preceding analysis shows that in
all the circumstances at least one of the effective potentials has a well
structure and the highest well ($V_{eff}^{-}$) governs the value of the
zero-point energy. Note that $s$ is equal to $q$ for $V_{eff}^{+}$, and $q+1$
for $V_{eff}^{-}$, with the additional possibility of $s=-q+1$ for $%
V_{eff}^{+}$ when $0<q<1$. Therefore, the demand for a common effective
eigenvalue implies that $n_{-}=n-1$, where $n=n_{+}$ is related to $\Psi _{+}
$ and $n_{-}$ is related to $\Psi _{-}$. The extra value for $s$ when $0<q<1$
(for $V_{eff}^{+}$) shows to be an unacceptable solution because it does not
provide an integer value for $n_{+}-n_{-}$. The solutions, including the
zero-mode ones, can now be written as

\begin{equation}
\Psi _{+}=N_{+}\xi ^{q}e_{\;}^{-\xi /2}\;L_{n}^{2q-1}\left( \xi \right) ,%
\hspace{0.25cm}{\rm (} q\ge 1 \hspace{0.25cm}{\rm for}\hspace{0.25cm}%
 n=0\rm {)}  \label{23a}
\end{equation}
\vspace{1cm}
\begin{equation}
\Psi _{-}=N_{-}\xi ^{q+1}\,e_{\;}^{-\xi /2}\;L_{n-1}^{2q+1}\left( \xi
\right) \;  \label{23b}
\end{equation}
\vspace{1cm}
\begin{equation}
E=\pm mc^{2}\sqrt{1-\left( \frac{q}{q+n}\right) ^{2}},\qquad
n=0,1,2,\ldots \label{23c}
\end{equation}

\noindent where we have used $L_{0}^{k}\left( \xi \right) =1$ and $%
L_{-1}^{k}\left( \xi \right) =0$ for all $k$. Note that $E\approx mc^{2}$
and $E_{eff}\approx E-mc^{2}$, is valid as long as $n\gg q$, thus the
nonrelativistic limit of the theory would be, in a limited sense, a regime
of large quantum numbers. On the other hand, in the regime of strong
coupling, \textit{i.e.}, for $q\gg 1$, one has $E_{eff}\approx
-mc^{2}/2+nmc^{2}/q$ and as the coupling becomes extremely strong the lowest
effective eigenvalues end up close to $-mc^{2}/2$, corresponding to Dirac
eigenvalues $E$, near zero. Now one sees clearly that the eigenvalues as
well as the eigenfunctions for a zero-energy solution, in contrast to what
is declared in Ref. \cite{cnt}, can be obtained as a limit case of a
nonzero-energy solution. Fig. \ref{Fig2} shows the behavior of the
positive-eigenenergies as a function of $q$.

The effective Kratzer potential leads to bounded solutions in the range $%
-mc^{2}/18\le E_{eff}<0$ when a critical potential ($V_{eff}^{+}$) is there
and $V_{eff}^{-}$ has a minimum value equal to $-mc^{2}/6$. For $q=1$ the
minimum value of $V_{eff}^{-}$ is $-mc^{2}/4$ and the effective eigenvalue
is greater than or equal to $-mc^{2}/8$. For $q>1$ the effective potential $%
V_{eff}^{\pm }$ has the minimum value $-mc^{2}q/2(q\mp 1)$ and the effective
eigenvalues are greater than or equal to $-mc^{2}q^{2}/2(q+1)^{2}$. In this
last case the fermion tends to avoid the origin more and more as $q$
increases due to higher and higher centrifugal barrier $\left[ \hbar
^{2}q(q\mp 1)/2mx^{2}\right] $, which in turn thrust the minimum of the
effective potential further (this shifting is proportional to ($q\mp 1$))
and higher (lower) for $V_{eff}^{+}$ ($V_{eff}^{-}$) in such away that $%
V_{eff}^{\pm }\rightarrow V_{eff}^{\mp }$.

\begin{figure}[!ht]
\begin{center}
\includegraphics[width=8cm, angle=270]{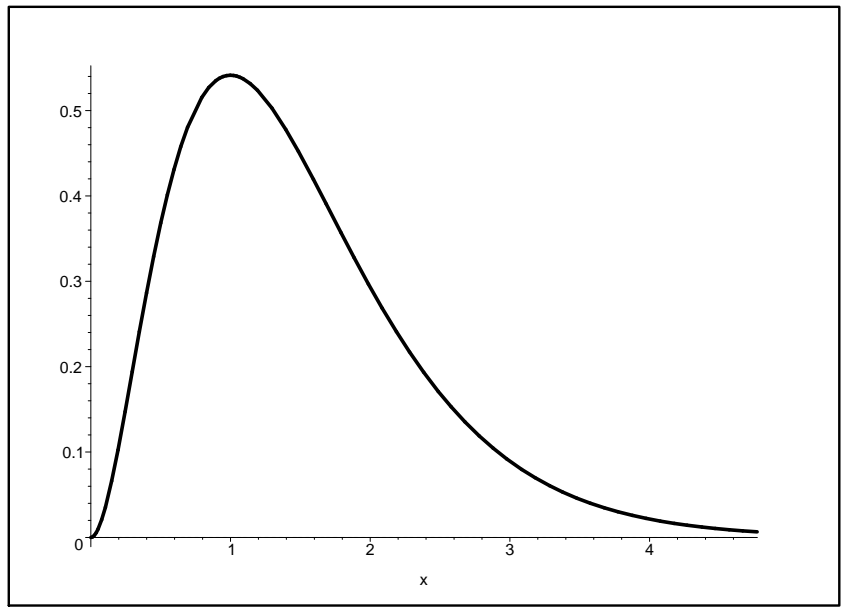}
\end{center}
\par
\vspace*{-0.1cm}
\caption{$|\Psi |^{2}=|\Psi _{+}|^{2}+|\Psi _{-}|^{2}$ as a function of $x$
corresponding to the ground state ($E=0$) for the potential $V=-\hbar cq/|x|$
with $q=1$ ($m=c=\hbar =1$).}
\label{Fig3}
\end{figure}

\begin{figure}[!ht]
\begin{center}
\includegraphics[width=8cm, angle=270]{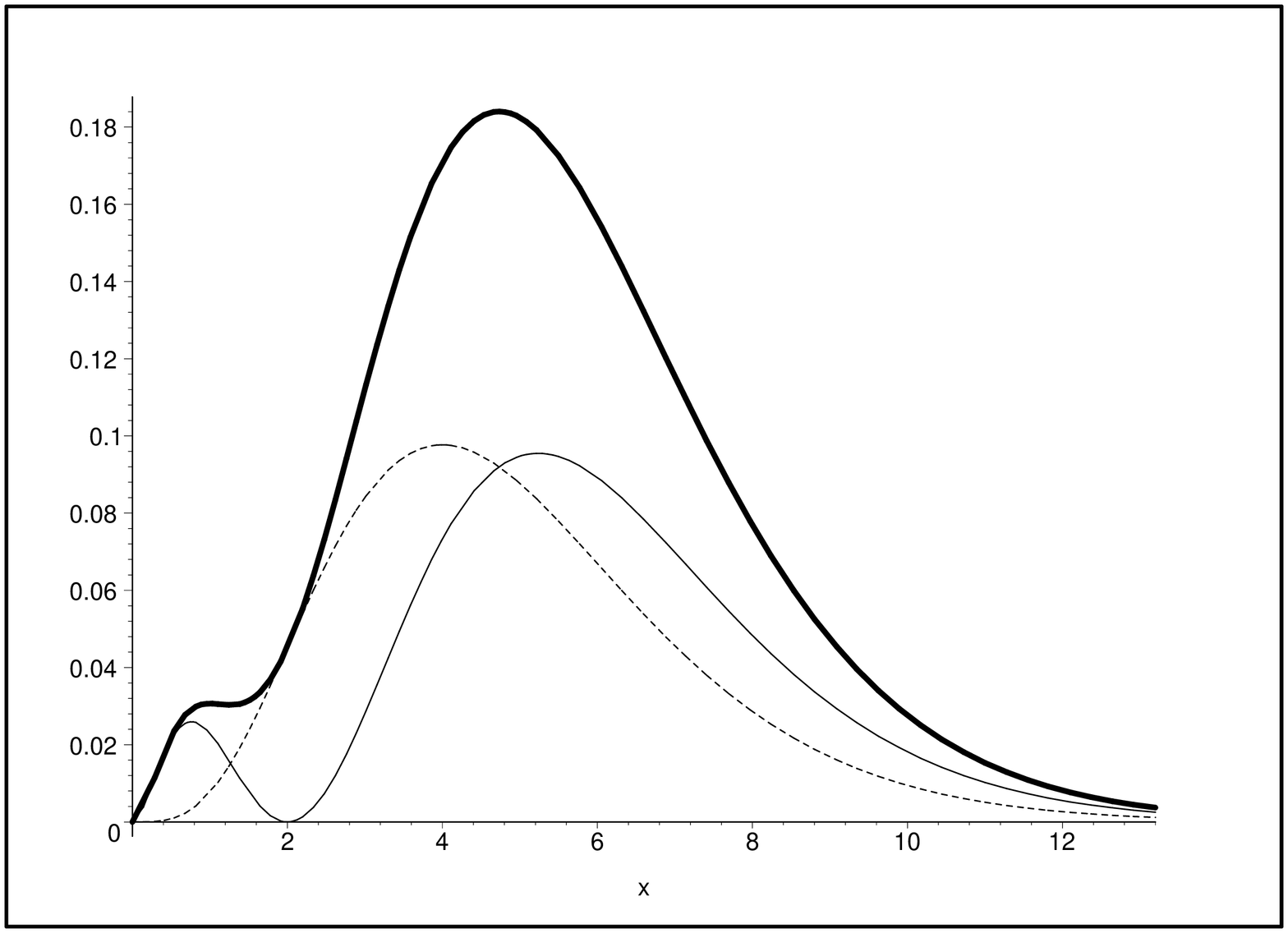}
\end{center}
\par
\vspace*{-0.1cm} \caption{$|\Psi _{+}|^{2}$ (full thin line),
$|\Psi _{-}|^{2}$ (dashed line) and $|\Psi |^{2}=|\Psi
_{+}|^{2}+|\Psi _{-}|^{2}$ (full thick line) as a function of
$x$, corresponding to the first-excited state ($E>0$) for the
potential $V=-\hbar cq/|x|$ with $q=1$ ($m=c=\hbar =1$).}
\label{Fig4}
\end{figure}

\begin{figure}[!ht]
\begin{center}
\includegraphics[width=8cm, angle=270]{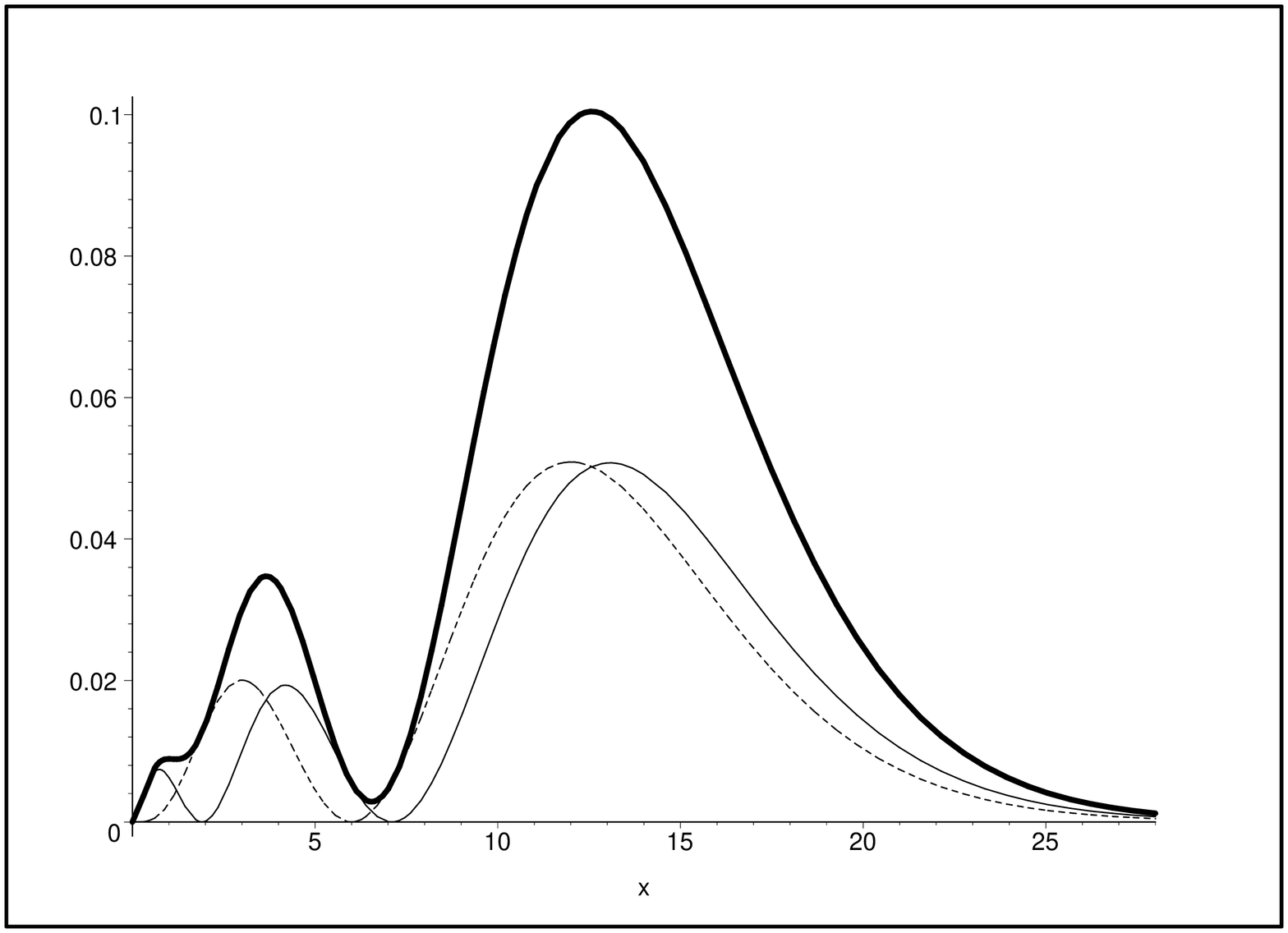}
\end{center}
\par
\vspace*{-0.1cm}
\caption{$|\Psi _{+}|^{2}$ (full thin line), $|\Psi _{-}|^{2}$ (dashed line)
and $|\Psi |^{2}=|\Psi _{+}|^{2}+|\Psi _{-}|^{2}$ (full thick line) as a
function of $x$, corresponding to the second-excited state ($E>0$) for the
potential $V=-\hbar cq/|x|$ with $q=1$ ($m=c=\hbar =1$).}
\label{Fig5}
\end{figure}

\begin{figure}[!ht]
\begin{center}
\includegraphics[width=8cm, angle=270]{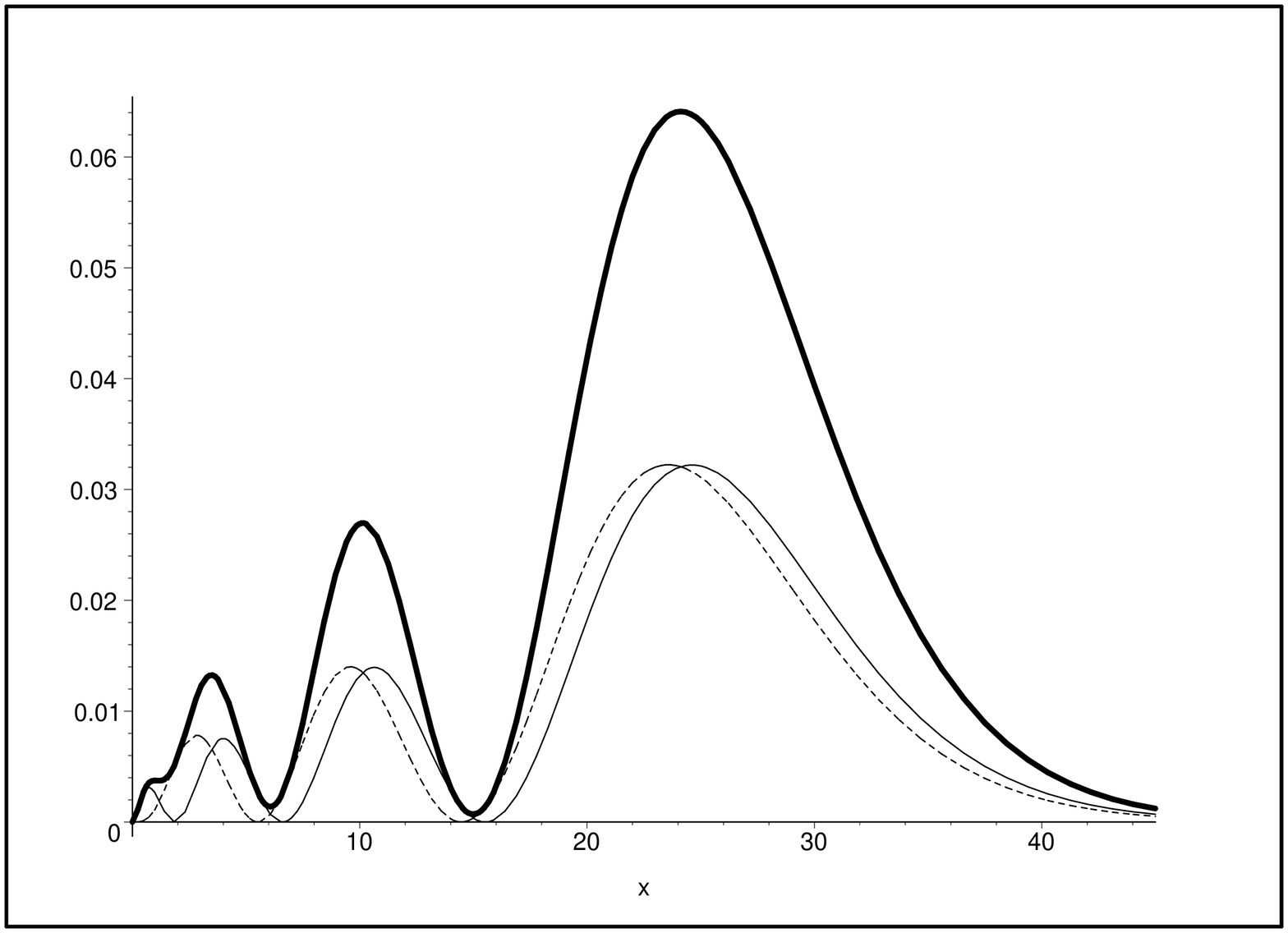}
\end{center}
\par
\vspace*{-0.1cm}
\caption{$|\Psi _{+}|^{2}$ (full thin line), $|\Psi _{-}|^{2}$ (dashed line)
and $|\Psi |^{2}=|\Psi _{+}|^{2}+|\Psi _{-}|^{2}$ (full thick line) as a
function of $x$, corresponding to the third-excited state ($E>0$) for the
potential $V=-\hbar cq/|x|$ with $q=1$ ($m=c=\hbar =1$).}
\label{Fig6}
\end{figure}

Fig. \ref{Fig3} illustrate the behavior of the probability position density,
$|\Psi |^{2}=|\Psi _{+}|^{2}+|\Psi _{-}|^{2}$, for the zero-mode solution
for $q=1$. Figs. \ref{Fig4}, \ref{Fig5} and \ref{Fig6} illustrate the
behavior of the upper and lower components of the Dirac spinor, $|\Psi
_{+}|^{2}$ and $|\Psi _{-}|^{2}$, and the probability position density, $%
|\Psi |^{2}=|\Psi _{+}|^{2}+|\Psi _{-}|^{2}$, for the positive-energy
solutions of the first-, second- and third-excited states for $q=1$. The
results for negative energies are the same as far as the charge conjugation
does $\Psi _{+}\rightarrow \Psi _{+}^{*}$ and $\Psi _{-}\rightarrow -\Psi
_{-}^{*}$ The relative normalization of $\Psi _{+}$ and $\Psi _{-}$ is
obtained by substituting the solutions directly into the original
first-order coupled equations (\ref{3}). The result is not pretty for $E\neq
0$ and we preferred to do it by numerical computation with a symbolic
algebra program. Comparison of these figures shows clearly that, except for
the zero-eigenmode ($n=0$), $|\Psi _{-}|$ is comparable in importance to $%
|\Psi _{+}|$ and $|\Psi _{-}|\rightarrow |\Psi _{+}|$ as $n$ increases,
except for a minor difference near the origin. It should not be forgotten,
though, that we have restricted our discussion to the positive half line.
For $x\leq 0$ the roles of $\Psi _{+}$ and $\Psi _{-}$ are reversed.

In conclusion, we have succeed in searching for Dirac bounded solutions for
the scalar potential $V=-\hbar cq/|x|$. The satisfactory completion of this
task for nonzero eigenenergies has been alleviated by the methodology of
effective potentials which has transmuted the question into Sturm-Liouville
problems with effective inversely linear plus inversely quadratic potentials
for both components of the Dirac spinor. A discrete and nondegenerate
spectrum has been found. For a strong enough potential a nondegenerate
zero-eigenmode corresponding to the ground state comes to the scene. Only in
the regime of extremely strong coupling there are\emph{\ }degenerate
solutions related to the zero-energy solutions, one of the eigenspinor
representing the fermion and the other representing the antifermion. Beyond
its intrinsic importance as a new solution for a fundamental equation in
physics, the problem analyzed in this paper presents unusual results.

\bigskip

\bigskip

{\small \noindent }\textbf{Acknowledgments}

The author wishes to thank M.B. Hott for useful discussions and for a
critical reading of the paper. This work was supported in part by means of
funds provided by CNPq and FAPESP.

\medskip \pagebreak

\end{document}